\documentstyle{mn} \input{epsf} \newif\ifAMStwofonts 

\def\lesssim{\mathrel{\hbox{\rlap{\hbox{\lower4pt\hbox{$\sim$}}}\hbox{$<$}}}}

\def\gtrsim{\mathrel{\hbox{\rlap{\hbox{\lower4pt\hbox{$\sim$}}}\hbox{$>$}}}}

\def\msun{$M_{\odot}$}

\def\teff{$T_{\rm eff}$~}

\def\lteff{$\log{T_{\rm eff}}$~}

\def\ll_lsun{$\log{\rm (L/L_{\odot})}$~}

\def\masa_msun{$M/ \rm M_{\odot}$~}

\def\m_mstar{$M/M_{*}$~}

\def\mh{${\rm M_H/M_*} $ }

\def\lmh{ $\log{{\rm (M_H/M_*)}}$ }

\def\lmhe{ $\log{ {\rm (M_{He}/M_*)} }$ }

\title[Asteroseismology  of G117-B15A]  {Time  dependent diffusion  in
pulsating white dwarf stars: Asteroseismology of G117-B15A}

\author[O.  G. Benvenuto,  A.  H. C\'orsico, L. G.   Althaus and A. M.
Serenelli]  {O.   G.    Benvenuto\thanks{Member  of  the  Carrera  del
Investigador    Cient\'{\i}fico,    Comisi\'on   de    Investigaciones
Cient\'{\i}ficas de la Provincia de Buenos Aires, Argentina.},
A.    H.    C\'orsico\thanks{Fellow  of   the   Consejo  Nacional   de
Investigaciones Cient\'{\i}ficas y T\'ecnicas (CONICET), Argentina.},
L.   G.    Althaus\thanks{Member  of  the   Carrera  del  Investigador
Cient\'{\i}fico y Tecnol\'ogico, CONICET.},
A.  M.  Serenelli\thanks{Fellow of  CONICET.}  \\ Facultad de Ciencias
Astron\'omicas  y Geof\'{\i}sicas, Universidad  Nacional de  La Plata,
Paseo  del   Bosque  S/N,   (1900)  La  Plata,   Argentina.\\  Emails:
obenvenuto,acorsico,althaus,serenell@fcaglp.fcaglp.unlp.edu.ar}

\pagerange{\pageref{firstpage}--\pageref{lastpage}}

\pubyear{2001}

\begin{document}

\maketitle

\label{firstpage}

\begin{abstract} 

We study the structural characteristic  of the variable DA white dwarf
G117B-15A  by applying the  methods of  asteroseismology.  For  such a
purpose, we  construct white  dwarf evolutionary models  considering a
detailed and  up - to -  date physical description as  well as several
processes  responsible for  the occurrence  of element  diffusion.  We
have considered  several thickness  for the outermost  hydrogen layer,
whereas  for the  inner  helium-, carbon-  and  oxygen-rich layers  we
considered realistic  profiles predicted by calculations  of the white
dwarf progenitor evolution.  The stellar masses we have analysed cover
the mass range of $0.50 \leq {\rm M}_{*}/M_{\odot} \leq 0.60$.

The evolution of each of  the considered model sequences were followed
down to very low effective  temperature; in particular, from 12500K on
we computed  the dipolar,  linear, adiabatic oscillations  with radial
order $k= 1, \cdots, 4$.  We find that asteroseismological results are
not univocal regarding mode  identification for the case of G117B-15A.
However,   our  asteroseismological   results   are  compatible   with
spectroscopical data only if the  observed periods of 215.2, 271.0 and
304.4 s are due to dipolar modes with $k=2, 3, 4$ respectively.

Our calculations  indicate that  the best fit  to the  observed period
pattern of G117B-15A corresponds to  a DA white dwarf structure with a
stellar  mass   of  0.525  \msun,   with  a  hydrogen   mass  fraction
\lmh$\gtrsim$-3.83  at an effective  temperature \teff$\approx$11800K.
The  value  of the  stellar  mass  is  consistent with  that  obtained
spectroscopically by Koester \& Allard.

\end{abstract}


\section{Introduction} \label{sec:intro}

Asteroseismology  has  become a  powerful  method  to disentangle  the
internal structure  and evolution  of stars by  means of the  study of
their oscillatory pattern.  This  technique, very sophisticated in the
case  of the Sun,  has also  undergone a  strong development  in other
stars,  in  particular  in   variable  white  dwarf  (WD)  stars  (see
e.g. Brown \& Gilliland 1994; Gautschy \& Saio 1995; 1996).

Pulsating WDs show multiperiodic luminosity variations in three ranges
of  effective  temperatures  (\teff)  corresponding to  the  currently
called DOV,  DBV and DAV (see,  e.g., the review by  Winget 1988).  Of
interest in this work are the DAVs (hydrogen - dominated atmospheres),
or ZZ  Ceti, that  pulsate in the  instability strip  corresponding to
11000 K $\leq$  \teff $\leq$ 13000 K.  The  periodicities in the light
curves of pulsating WDs are  naturally explained in terms of nonradial
g-modes  of  low  harmonic  degree  ($\ell \leq  2$),  driven  by  the
``$\kappa$ mechanism'' working in  a partial ionization region near to
the stellar  surface (Dolez  \& Vauclair 1981,  Winget et  al.  1982).
The periods ($P$) are found within a  range of $100 s \leq P \leq 1200
s$ and photometric amplitudes reach up to 0.30 magnitudes.

Most  asteroseismological studies  performed on  WDs to  date  rely on
stellar models  constructed under some simplifying  hypotheses. One of
the most relevant  is related to the profile  of the internal chemical
composition in the interface  zones.  In such regions, the equilibrium
diffusion in the trace  element approximation has been widely employed
to  infer  the profile  of  the  chemical  distribution (see  Tassoul,
Fontaine \&  Winget 1990).  The  main motivation for  considering this
approximation  is to  avoid  the solution  of  time dependent  element
diffusion as  the WD evolves. In  the frame of  such an approximation,
the profile  of the  interface region is  very simple:  its functional
form is a power law. The  transition zone is separated into two parts:
an  upper one  in  which one  element  is dominant  and  the other  is
considered as  a trace  and a lower  region in  which the role  of the
respective  elements  is  reversed.    Because  these  two  power  law
solutions are matched for  fulfilling the condition of conservation of
mass of each element, a discontinuity in the derivative occurs just at
the matching point. The exponent of the power law solution is directly
related to the  state of ionization of the  stellar plasma.  Thus, the
structure of the  whole interface zone, in the  frame of this standard
treatment,  can  be  modified  only  if  the  plasma  suffers  from  a
modification  in  the state  of  ionization  as  a result  of  stellar
evolution.  Calculations  of asteroseismology of  WDs in the  frame of
such a  standard treatment for  the chemical interfaces are  those of,
e.g., Bradley  (1996, 1998, 2001), Bradley \&  Winget (1994), Brassard
et al.  (1991,  1992ab), Fontaine et al. (1992),  Montgomery \& Winget
(1999),  Metcalfe,  Nather  \&  Winget  (2000),  Metcalfe,  Winget  \&
Charbonneau (2001), Montgomery, Metcalfe \& Winget (2001).

From  an  evolutionary  point  of  view, the  shape  of  the  chemical
interfaces  may  not  be  critical\footnote{An obvious  and  important
exception is when the tail of chemical profiles are subject to nuclear
burning.}, but in pulsational studies  they provide a non - negligible
contribution   to   the   shape    of   the   Ledoux   term   of   the
Brunt-V\"ais\"al\"a frequency (Brassard et  al.  1991).  Thus, we can,
in  principle,  expect differences  between  studies with  equilibrium
diffusion and those in which  other more physically sound treatment is
performed. This  is a  very important point  to be made  in connection
with the aim of the present work.

Since the  pioneering work of Iben  \& MacDonald (1985),  we know that
element diffusion modifies  the chemical abundance distribution within
a WD star even during evolutionary stages corresponding to the ZZ Ceti
domain (see Iben  \& MacDonald 1985, particularly their  Fig.  4). Few
calculations exist in the literature  in which the evolution of WDs is
addressed  in  a  self  consistent  way with  time  dependent  element
diffusion. Amongst them, we mention the study of MacDonald, Hernanz \&
Jos\'e  (1998)  aimed  at   studying  the  carbon  pollution  in  cool
WDs. Also, Dehner  \& Kawaler (1995) used non  - equilibrium diffusion
together  with  evolutionary calculations  to  study  WDs with  helium
envelopes.   Finally,  Althaus,  Serenelli  \& Benvenuto  (2001)  have
recently shown that diffusion  induces the occurrence of thermonuclear
flashes in  helium - core WDs,  causing the evolution of  such kind of
WDs to  occur on timescales significatively shorter  than predicted by
models  without diffusion.   This has  been particularly  important in
solving the discrepancy  (Van Kerkwijk et al.  2000)  about the age of
binary systems containing a millisecond pulsar and a helium WD.

As far  as we  are aware, the  only work  aimed at exploring  the role
played by diffusion in the period and its rate of change for $g$-modes
in  pulsating  DA  WDs is  that  of  C\'orsico  et al.   (2001b).   In
particular, the authors found that the differences in the shape of the
chemical profiles at the interface zones induce appreciable changes in
the periods, as  compared to the case of  equilibrium diffusion in the
trace element  approximation.  Also,  there are noticeable  changes in
the period derivative,  which are due in part to  the evolution of the
chemical  profile during  the cooling  of the  WD across  the  ZZ Ceti
instability strip.

Since  sometime ago  there have  been several  works available  in the
literature  in which  the observed  period structure  of  a particular
object is fitted to theoretical  predictions. Such works show that, in
principle, information about the stellar mass and the stratified outer
layer  structure can  be inferred  (see,  e.g., Winget  et al.   1991;
Fontaine et al. 1992; Pfeiffer et al.  1996; Bradley \& Kleinman 1997;
Bradley 1998,  2001; Bradley  \& Winget 1994).   As these  studies are
based on  a simplified treatment of  diffusion, we believe  that it is
worth revisiting this problem on  the basis of more detailed models of
WD structure. In particular, in view of the results found in C\'orsico
et al. (2001b), we expect to find differences in the period fitting to
a  particular object,  as compared  with  the situation  in which  the
standard treatment is used. The only  way to find how important can be
such differences is by performing a detailed asteroseismological study
choosing a well studied object.

We consider G117-B15A  as an optimal target for  our study.  G117-B15A
is an otherwise typical DA WD, the variability of which was discovered
by Mc Graw  \& Robinson (1976) and, since then,  it has been monitored
continuously.  The mass and in particular the effective temperature of
this   star  have   been   the  subject   of  numerous   spectroscopic
redeterminations (see Gautschy, Ludwig \& Freytag 1996 for a summary),
in particular  values of  0.59 \msun and  11620 K,  respectively, have
been derived  by Bergeron et  al.  (1995).  More recently,  Koester \&
Allard (2000) (hereafter KA) have suggested a somewhat lower value for
the  mass  of  0.53  \msun  and  a  higher  effective  temperature  of
\teff=$11900 \pm 140$ K.

G117-B15A has periods of oscillation of 215.2, 271 and 304.4 s (Kepler
et al.  1982).  Notably, for the  215.2 s mode it has been possible to
find a  value of its temporal  derivative (Kepler et al.   2000) to be
$\dot{P}= (2.3  \pm 1.4) \times 10^{-15}$  s s$^{-1}$.  Interestingly,
the 215.2  s mode  present in G117-B15A  is the most  stable optically
registered oscillation with a stability comparable to that of the most
stable millisecond  pulsars.  As G117-B15A is a  well know oscillator,
it has motivated  the interest of several researchers  in its internal
structure.  For example, Bradley (1998)  has found the best fitting to
the period pattern  with a model of $\approx  0.6$ \msun and depending
on the  identification of the modes  (see below for  an explanation of
its  meaning),  the favoured  value  for  the  hydrogen envelope  mass
fraction \mh is $-7  \lesssim \log{{\rm (M_H/M_*)}}\lesssim -6$ or $-5
\lesssim  \log{{\rm  (M_H/M_*)}}\lesssim   -4$,  and  \lmhe=-2.   More
recently, C\'orsico  et al.   (2001a) have found,  on the basis  of WD
models in which diffusion is neglected, the best fitting to the period
pattern with  a 0.55 \msun WD  model with a carbon  - oxygen interior,
\lmhe=-2 and \lmh=-4 as  predicted by stellar evolution. Notably, such
a fitting is  in nice agreement with one of  those proposed by Bradley
(1998).

It  is the  aim  of  this work  to  perform a  fitting  to the  period
structure present in G117-B15A  by computing the non-radial eigenmodes
in the  frame of linear and adiabatic  approximation, and evolutionary
WD  models  in which  time  dependent  element  diffusion is  properly
accounted for.   About the way we  shall perform such  a fitting, some
words are in order.  In handling  models like those we shall employ it
is quite  obvious that we are  not in a position  to employ techniques
like  those  used  in the  so  -  called  genetic algorithm  (see  its
application for the  case of the DBV GD358 in  Metcalfe et al.  2000).
In the  context of ZZ Ceti stars  a similar approach though  to a less
extent has  been repeatedly  employed by numerous  investigations (e.g
the complete and detailed study  performed by Bradley 1998, 2001).  Up
to  now,  the  shape  and  thickness  of  the  relevant  chemical
interfaces are usually  treated as a free parameter.   However, a more
physically sound  treatment can be  performed when account is  made of
time dependent  element diffusion in evolutionary models  of WDs. This
aspect of WD evolution is one of the most important when an attempt is
made  to  compare  observations  with  theoretical  expectations  from
pulsating WDs.  Unfortunately,  there are currently some uncertainties
(e.g., the treatment  of convection, the rate of  the critical nuclear
reaction ${^{12}}{\rm C}(\alpha,\gamma){^{16}}{\rm O}$, wind mass loss
episodes,  etc.) that  prevent us,  provided a  mass value  and pre-WD
composition, from predicting a whole definite internal structure. This
is especially  true for \mh.  Thus,  in view of these  facts, we shall
consider the  value of  \mh as a  free parameter. We  assumed \lmhe=-2
throughout this paper.

The  remainder  of the  paper  is  organized  as follows:  In  Section
\ref{sec:comput} we discuss the  physical ingredients we have employed
together  with the  computational strategy  we have  employed. Section
\ref{sec:evolution}  is  devoted  to  presenting the  details  of  the
evolutionary  models  we  have  constructed  down  to  the  conditions
relevant   for   the   WD    we   are   investigating.    In   section
\ref{sec:asteroseis}  we present  the asteroseismological  analysis we
performed   for   the   particular   case   of   G117B-15A.    Section
\ref{sec:dicuss} is devoted  to the discussion of the  results we have
found and finally in Section \ref{sec:conclus} we give some concluding
remarks and insights for future investigations.

\section{Details of the Computations} \label{sec:comput}

For this work we have  employed the same code and physical ingredients
as in C\'orsico \& Benvenuto (2002) and C\'orsico et al. (2001ab). In
particular  for the  diffusion processes  we  considered gravitational
settling,  chemical and  thermal diffusion as in Althaus  \& Benvenuto
(2000).

In our  code, evolutionary and pulsational  calculations are performed
in a fairly automatic way. After selecting a starting stellar model we
choose an interval  in $P$ and \teff.  The  evolutionary code computes
the model cooling until the hot edge of the \teff-interval is reached.
Then, the  program calls the  set of pulsation routines  beginning the
scan  for  modes.   When  a  mode  is found,  the  code  generates  an
approximate solution which is  iteratively improved to convergence (of
the  eigenvalue and  the eigenmode  simultaneously) and  stored.  This
procedure is repeated until the period interval is covered.  Then, the
evolutionary code generates the next stellar model and calls pulsation
routines again.  Now, the previously stored modes are taken as initial
approximation to  the modes  of the subsequent  model and  iterated to
convergence.   Such  a procedure  is  automatically  repeated for  all
evolutionary  models   inside  the  chosen  \teff   -  interval.   The
computational strategy  described above has  been successfully applied
in fitting the observed periods  of G117-B15A to impose constraints on
the mass of axions (C\'orsico et al.  2001a) in the frame of WD models
less detailed than those we shall employ in the present work. We refer
the  reader  to  C\'orsico   \&  Benvenuto  (2002)  and  C\'orsico  et
al. (2001ab) and references therein for further details.  Let us quote
that most  of our evolutionary models  have been divided  in more than
2000  meshpoints,  whereas  for  mode calculations  we  employed  5000
meshpoints.

In order  to start  our evolutionary calculation  we have  employed an
artificial heating technique detailed  in Althaus \& Benvenuto (2000).
Let us  quote that in this  technique, the stellar model  is forced to
undergo some unphysical evolution  by introducing an artificial energy
release  that, when  the  star  gets bright  enough,  is switched  off
smoothly.  Then, the star relaxes to the physical cooling branch after
few tens of models.

In employing  this artificial  technique, we must  be very  careful in
avoiding generating an artificial  chemical profile. Thus, in order to
verify that the  star not only relaxes to  the physical cooling branch
but  {\it  simultaneously}  relaxes  to a  correct  internal  chemical
profile, we have employed, for a fixed \lmh value, different shapes of
the  initial hydrogen  and helium  profiles.   We found  that, if  the
initial model  is far away enough  from the evolutionary  stage we are
interested in, the fine details  of the initial profile are irrelevant
because  diffusion largely  evolves it  to a  well defined  one.  As a
matter of fact, the structure and the chemical profiles are physically
plausible far before the  star attains the effective temperature range
relevant for G117-B15A.

A better approach for constructing pulsating WD models would obviously
be to  start computing  evolutionary models from  much earlier  in the
star history. This possibility, better  from a physical point of view,
again  would  require an  enormous  amount  of  computing time  for  a
definite sequence. Thus  trying to fit a period pattern  in such a way
is, in  our opinion, far beyond  what seems reasonable  at the present
state of the art.  In any case, it would be interesting to explore the
pulsational properties  of models  resulting from a  full evolutionary
calculation.  We plan to present such a study in a future paper.

To  be specific,  we  have  started the  calculations  with models  of
\ll_lsun$\approx$ 2.05, \lteff$\approx$4.46  and masses of 0.50, 0.55,
and  0.60  \msun   (in  performing  the  fitting  of   the  models  to
observations we  have also computed supplementary models  of 0.525 and
0.5375 \msun).   The mass fraction embraced by  the outermost hydrogen
layer has been assumed in a way that, after evolving along the cooling
branch down  to \teff=12500 K the  values of \lmh present  in the star
are  those given  in  Table 1.   At  the starting  conditions we  have
assumed the  Salaris et al.   (1997) core composition, whilst  for the
outer layers the CNO abundances were those corresponding to the pre-WD
evolution. In  order to get different  values of \mh,  we have started
from the  model with  the largest \mh  and simply replaced  $^{1}$H by
$^{4}$He at the base of the  hydrogen envelope.  Let us quote that the
initial model  has internal temperatures  low enough that  prevent the
occurrence of nuclear reactions up to the moment at which the WD model
has relaxed  to the  physical branch.  We  should mention  that models
become  physically  adequate  (after  the unphysical  transitory)  for
describing  a WD  at \ll_lsun$\approx$  1.4,  \lteff$\approx$4.83, far
from the conditions relevant for DAV  WDs. Also, the values of \lmh at
the instability strip are somewhat  lower than those included in Table
1  simply  because  nuclear  reactions  are still  operating  at  such
evolutionary stages.  However this effect  is noticeable only  for the
case of  the thickest hydrogen  envelopes on the most  massive objects
considered here.

\begin{center}
\centerline{Table 1. {\small Values of \lmh for our computed models}}  
{{\small at \teff=12,500 K}}
\begin{tabular}{ccccc}
\hline \hline $0.50 M_{\odot}$ &  $ 0.55 M_{\odot}$ & $0.60 M_{\odot}$
& $0.525 M_{\odot}$  & $0.5375 M_{\odot}$ \\ \hline  -3.815 & -3.862 &
-3.941 & -3.831 & -3.843 \\ -4.193 & -4.224 & -4.249 & -4.198 & \ldots
\\ -4.679  & -4.684 & -4.692  & \ldots &  \ldots \\ -5.175 &  -5.177 &
-5.180 & \ldots & \ldots \\ -5.671 & -5.671 & -5.672 & \ldots & \ldots
\\ -6.160  & -6.158 & -6.156  & -6.159 &  \ldots \\ -6.640 &  -6.633 &
-6.700 & -6.634 & \ldots \\ -7.071 & -7.047 & -7.028 & \ldots & \ldots
\\ -7.423 & -7.385 & -7.349 & \ldots & \ldots \\ \hline
\end{tabular}
\end{center}

To our notice, this is the first opportunity in which carbon - oxygen,
intermediate mass WD models with a set of values in \mh are evolved in
the frame  of time dependent  element diffusion.  Prior to  this work,
the kind of calculations carried  out here were performed only for the
values of \mh predicted by stellar evolution theory (Iben \& MacDonald
1985; 1986).

\section{Evolutionary Results} \label{sec:evolution}

For each  initial model,  we have computed  its evolution down  to the
conditions corresponding  to the DAV instability strip.   In doing so,
we computed about ten thousand  models for each sequence.  Perhaps the
most relevant  characteristic of the models  for the kind  of study we
are  performing  here  is  the  internal  chemical  profile.   Results
representative for  the case  of the models  corresponding to  the DAV
instability strip are  presented in Fig. 1.   In this figure we
depict the  internal chemical  profiles for a  0.55 \msun WD  model at
\lteff $\approx$  4.096 for  \lmh=-3.862.  The adopted  effective 
temperature  is slightly
higher  than  the   last  estimation  for  G117B-15A  (\lteff$\approx$
4.076).   Thus,  these  profiles  should  be  representative  of  the
conditions expected for the interior  of G117B-15A in the frame of our
set of models.

Let  us discuss  the main  characteristics of  these  profiles. Before
discussing  them, we  have to  warn the  reader that  the way  we have
modified  the value of  \mh is  ad-hoc. In  fact, it  is not  clear at
present  if  there  exists,  for   a  fixed  stellar  mass  value,  an
evolutionary mechanism capable to produce a distribution of \mh values
differing each other by orders  of magnitude. In addition, the outward
tail of  the distribution of  initial heavy element abundances  may be
different from the results we  shall present below. However, we should
remark that  the profile  of the light  elements present in  the outer
parts of the  models should be essentially correct.   

With regard to the chemical profiles  of our models we want to mention
that for  $\log{\rm(1-M_{r}/M_{*})} \gtrsim -1.6$ the  carbon - oxygen
distribution is that predicted by pre-WD evolutionary calculations and
it is the same for all of our evolutionary sequences.  However this is
{\it not}  the case  for the chemical  structure of the  outer layers,
which  are, in  turn, crucial  for the  pulsational properties  of the
models. The main  variations are related to the  whole distribution of
$^{1}$H,  $^{3}$He  and  the  outwards  tail  of  $^{4}$He.   While  a
variation in the assumed value of \mh obviously implies a modification
in the profiles,  we find a very noticeable variation  in the slope of
the profiles  for the isotopes present  in the outer layers  of our WD
models.   In the  model with  \lmh=-3.815 it  is remarkable  the large
slope  in the tail  of the  hydrogen distribution.   In spite  of this
steep  profile, degeneracy  prevents hydrogen  from  diffusing further
inwards.  In models  with a smaller hydrogen content,  the tail of the
hydrogen distribution  is smoother.  Note  also that the shape  of the
tail  of  the hydrogen  distribution  at  the evolutionary  conditions
relevant for DAV  WDs is strongly dependent on the value of the
assumed \mh.

Regarding the tail  of the $^{4}$He distribution, this  is largely due
to  gravitational settling  that  forces $^{4}$He  to  sink down.   In
particular  our models  also  present  a zone  in  which the  chemical
composition is  largely dominated by  helium. Also, it is  of interest
the distribution  of $^{3}$He. It  is clearly noticeable that  most of
the  $^{3}$He  is located  in  stellar  layers  at which  $^{1}$H  and
$^{4}$He are present.  This is  a consequence of the interplay between
nuclear  reactions  that form  $^{3}$He  by  means  of the  first  two
reactions of the PPI chain (Clayton 1968)\\

\noindent ${^1}{\rm H} + {^1}{\rm H} \rightarrow {^2}{\rm H} + e^{+} +
\nu$\\
 
\noindent ${^2}{\rm H} + {^1}{\rm H} \rightarrow {^3}{\rm He}$\\

and gravitational settling. Notice that the outer tail of the $^{3}$He
distribution is similar to the  corresponding to $^{4}$He which is, in
part, due to the operation of the reaction

\noindent ${^3}{\rm He} + {^3}{\rm He} \rightarrow {^4}{\rm He} + 2 \; 
{^1}{\rm H}$.

As the  chemical profile of the  hydrogen in surface layers  is one of
the key ingredients in determining the  g-modes of DAV WDs, we show in
Fig. 2 the base of the  hydrogen envelopes for our suite of 0.55 \msun
models  at the  ZZ Ceti  domain.  In  addition, we  show  the profiles
predicted by  the standard treatment  of equilibrium diffusion  in the
trace element approximation for the  same stellar mass and \mh values.
It is  clear that time  dependent element diffusion  predicts hydrogen
profiles that are somewhat different  from those given by the standard
treatment.  In  the standard  treatment, the slope  of the  profile is
determined solely by  the state of ionization of  the plasma (see Eqs.
26 - 29 of Tassoul et al.  1990).  Thus, if the state of ionization is
the same,  the slope should remain essentially  unaltered with further
evolution. We  have verified  that, even in  the case of  the thinnest
hydrogen  envelopes  we have  considered,  the  base  of the  hydrogen
envelope is located at layers with temperatures above $10^{6}$K, which
ensures complete  ionization for  hydrogen and helium  and this  is so
even  at   the  effective   temperatures  corresponding  to   the  DAV
instability strip. Also, notice that the sudden change of the slope at
the  matching point  of the  two solutions  of the  standard treatment
(denoted by circles along the  curves) has a direct consequence on the
enhancement of  the trapping  of some modes  in the  hydrogen envelope
(see, e.g., Brassard  et al.  1992a, Bradley 1996).  Thus, our results
show the  necessity of allowing for  a more detailed  treatment of the
chemical evolution at performing asteroseismology of WDs.

Now,  let  us discuss  the  impact of  our  chemical  profiles on  the
Brunt-V\"ais\"al\"a   frequency,  a   key  quantity   entering   as  a
coefficient in the equations of non-radial pulsations. This is defined
as (Unno et al. 1989)

\begin{equation}
N^{2} = g\  \left(\frac{1}{\Gamma_{1}}\ \frac{d\ln P}{dr} - \frac{d\ln
\rho}{dr} \right)
\end{equation}

\noindent where  all the symbols have  their usual meaning\footnote{In
this Section $P$ stands for the  pressure but in the rest of the paper
it stands for the period  of oscillation.} and $\Gamma_1$ is the first
adiabatic exponent.   However, in order  to avoid numerical  noise, we
use the expression given by Brassard et al.  (1991):

\begin{equation}
N^{2}=  {{g^{2}  \rho}\over{P}} {{\chi_{T}}\over{\chi_{\rho}}}  \left[
\nabla_{ad} - \nabla + B \right].
\end{equation}

The Ledoux term $B$, for the case of a M-component plasma, is given by

\begin{equation}
B=     -     {{1}\over{\chi_{T}}}    \sum_{i=1}^{M-1}     \chi_{X_{i}}
{{d\ln{X_{i}}}\over{d\ln{P}}}.
\end{equation}

where

\noindent    $$\chi_{\rho}=\left(   {{\partial   \ln{P}}\over{\partial
\ln{\rho}}} \right)_{T,\{X_{i}\}}$$
 
\noindent $$\chi_{T}=\left(  {{\partial \ln{P}}\over{\partial \ln{T}}}
\right)_{\rho,\{X_{i}\}}$$

\noindent   $$\chi_{X_{i}}=\left(   {{\partial   \ln{P}}\over{\partial
\ln{X_{i}}}} \right)_{\rho,T,\{X_{j\neq i}\}}.$$

In  Fig.  3 we  show the  squared Brunt-V\"ais\"al\"a frequency  and the
Ledoux term for models of 0.55 \msun with \lteff$\approx$4.096 and all
the  considered  values  of  \mh.   We  also  show  in  an  inset  the
Brunt-V\"ais\"al\"a frequency  at the hydrogen -  helium interface for
the  case   of  the  thickest  hydrogen  envelope   according  to  our
computations and to the prediction of equilibrium diffusion (thick and
thin lines, respectively). In order  to show the correspondence of the
structure of  these functions with  the internal chemical  profiles of
the  star  we have  also  included  the  curves corresponding  to  the
abundances of $^{4}$He and $^{12}$C.

The Ledoux term $B$ has a structure that remains largely invariant for
the stellar core under changes in  \mh. From the centre outwards, the three
peaks are  due to steep slopes  in the carbon profile.   The first two
are due to the structure of  the carbon - oxygen interface whereas the
other is  due to the  helium - carbon  interface.  In contrast  to the
internal  behaviour of the  Ledoux term,  the other  peak, due  to the
hydrogen  -  helium  transition,  is  largely modified,  not  only  in
position but also in its height. Notice that the lower \mh the thinner
the  peak (see  Brassard et  al.   1991 and  Tassoul et  al. 1990  for
comparison  of the  shape of  the $B$  term and  Bradley 1996  for the
profile of the Brunt-V\"ais\"al\"a frequency).

It is worthwhile to comment on  the fact that because of our realistic
chemical profiles,  we would  expect to find  differences in  the mode
trapping properties as compared with previous studies.  This important
issue surely deserves a careful exploration, which will be reported in
a subsequent communication.

\section{Asteroseismological Results} \label{sec:asteroseis}

An  important  aspect for  the  present  work  is the  so-called  mode
identification.   That is, the  identification of  the $\ell$  and $k$
values  corresponding  to  each  observed  period.   In  the  case  of
G117-B15A, the  results of Robinson  et al.  (1995) indicate  that the
215.2 s  period corresponds to  a dipolar mode  ($\ell=1$).  Following
the work  by Bradley (1998) we  shall assume that the  other two modes
cited  above are  also dipolar  (Brassard  et al.   1993; Fontaine  \&
Brassard 1994), and  also that the other periodicities  present in the
light curve of the star  are not associated with actual eigenmodes but
are actually  due to non-linear  effects in the envelope  (Brassard et
al.  1993).  With regard to the radial order of the modes, there exist
two  possible identifications: following  Clemens (1994)  the observed
periods  are dipolar modes  with $k=2$  (215.2 s),  $k=3$ (271  s) and
$k=4$ (304.4  s).  On the  other hand, Fontaine, Brassard  \& Wesemael
(1994) identified these  modes as dipolar with $k=1$  (215.2 s), $k=2$
(271 s) and  $k=3$ (304.4 s).  According to  these facts, in computing
the  g-modes of  our  WD  models, we  have  considered, as  previously
mentioned, only  $\ell=1$ and radial  order $k= 1, \cdots,  4$. Higher
radial orders correspond, for our stellar mass values, to periods long
in excess compared to those observed and thus are not considered here.

In  order to  look  for a  fit  to observations  we  shall consider  a
function $\psi$ defined as

\begin{equation}
\psi=  \log{  \sum_{i=1}^{3}  \left[1 -  {{P_{i}^{C}}\over{P_{i}^{O}}}
\right]^{2}},
\end{equation}

\noindent where $P_{i}^{O}$ and  $P_{i}^{C}$ are the i-th observed and
calculated   periods  respectively.   For   the  case   of  G117B-15A,
$P_{i}^{O}$= 215.2, 271  and 304.4 s.  As a  result of the pulsational
calculations  we have the  values of  the periods  for $k=  1, \cdots,
4$. Then, assuming the  two reasonable possible identifications of the
observed  modes (i.e.   $P_{i}^{O}$  corresponding to  $k=  1, 2,  3$,
hereafter case {\bf A}, and $P_{i}^{O}$ corresponding to $k= 2, 3, 4$,
hereafter case {\bf B}) we can compute the function $\psi$ in terms of
the  effective temperature  of the  star  for each  of the  considered
sequences. Obviously  in the case  that the fitting were  exact, $\psi
\rightarrow -\infty$.  Thus the criterium we shall assume in selecting
the best  model for representing G117B-15A  is simply to  look for the
minima of the $\psi$ function considering the whole set of models.  In
such  a way  we  shall be  able  to determine  not  only the  internal
structure    and   mass    of   G117B-15A,    but   also    its   {\it
asteroseismological} effective temperature.

In Fig.  4  we show $\psi$ as a function  of the effective temperature
for the case of the set of  models with 0.50 \msun. In this figure the
left panel labeled as {\bf A} corresponds to the identification of the
observed  modes with  $k= 1,  2, 3$  whereas the  right panel  {\bf B}
corresponds to  the other identification  of modes we  shall consider:
$k= 2, 3, 4$ modes.  As it has been found by Bradley (1998), for cases
{\bf A}  and {\bf B}  the best fitting  to observations are  found for
models with values of \mh very different.  While assuming case {\bf A}
we find the best fit with rather thin hydrogen envelopes (\lmh= -6.16)
and in  the case {\bf  B} the best  fit is found  for the case  of the
thickest considered hydrogen envelope:  \lmh= -3.81.  It is remarkable
that  the  minimum  found  for  case  {\bf A}  is  a  very  deep  one,
corresponding     to    a     mass    and     effective    temperature
(\teff$\approx$11400K)  rather lower than  those corresponding  to the
last determination of these parameters for G117B-15A.

If we  increase the stellar mass  to 0.55 \msun, the  behaviour of the
function $\psi$ is largely changed (see Fig.  5). Despite that in case
{\bf A} the best fitting is for \lmh= -6.63, it corresponds however to
an effective  temperatures far  lower than that  suggested by  KA (and
outside the DAV instability strip).   By contrast, in case {\bf B} the
best fit  is again  provided by the  model with the  thickest hydrogen
envelope, but  again it occurs at effective  temperatures somewhat too
low.

Finally, in Fig.  6 we show  the behaviour of function $\psi$ for the
case of the  models constructed assuming a stellar  mass value of 0.60
\msun. It is clear that in both cases of mode identifications the fits
are far poorer than those for  the other, lower mass values. Thus, the
results  presented  up  to   this  point  strongly  suggest  that  the
asteroseismological mass of G117B-15A  should be in between 0.50 \msun
and 0.55  \msun. Moreover, the only  viable mass for the  case {\bf A}
seems to be 0.50 \msun, because larger mass values give rise to poorer
fits.

Let  us  analyse  the  results  corresponding to  case  {\bf  B}  more
carefully,  centering  our  attention  in  models  with  the  thickest
hydrogen envelopes. From  Fig. 4b, it can be noticed  that in the case
of  0.50 \msun  model,  the  fit is  better  for increasing  effective
temperatures. However, notice that (Fig.  5b) for the 0.55 \msun model
the  situation  is just  the  opposite:  the  fit improves  for  lower
effective  temperatures. 
Guided  by this  fact, it  can be  expected a
minimum in $\psi$ for some  intermediate value of the stellar mass. 
In order to explore this possibility we computed 
models with  0.525 \msun restricting ourselves to values
of \mh  similar to those  that provided good  fits (see Table  1). For
this mass value,  the identification of modes in case  {\bf A} shows a
good fit  for \lmh=  -6.63 at \teff$\approx$11400  while case  {\bf B}
presents  an  interesting  minimum   at  stellar  mass  and  effective
temperature values  very similar  to those determined  by KA  for this
object by spectroscopic analysis.
Finally, in order to perform a finer analysis, we compute a sequence
of models with  0.5375 \msun and a thick hydrogen envelope (see Table  1). 
The results for  the whole  set of  masses and  thickest hydrogen
envelopes for  case {\bf B} is shown  in Fig. 7.  The  labels of each
curve correspond to the values of \lmh presented in Table 1.
  
The results presented in this  Section indicate that, if we expect the
asteroseismological masses and effective temperatures to be similar to
those determined  spectroscopically, the  best fit is  found in the
frame of case {\bf B} for a model with a stellar mass of 0.525 \msun.
For the sake of completeness, we  also show in Fig.  8 the periods of
oscillation for radial order $k=2,3,4$ for the case of models with the
thickest hydrogen  envelope for each  considered value of  the stellar
mass.

\section{Discussion} \label{sec:dicuss}

From the results presented in  the preceding Section, we conclude that
the best fit to observations is found with stellar mass and \mh values
that are very sensitive to  the adopted mode identification.  This has
been previously found  by Bradley (1998). If we  restrict ourselves to
analyse the case of G117B-15A only in the frame of asteroseismological
models we find no way to be conclusive about the actual values for its
stellar  mass,  \mh,  and   effective  temperature.   However,  if  we
incorporate the best available spectroscopic determinations of stellar
mass and effective temperatures to this analysis and expect them to be
compatible  with the asteroseismological  results, we
can discriminate between  the identification of modes and  also on the
other values for the characteristics of the models.

In our opinion, the  best fit found in the frame of  case {\bf A} mode
identification,  though being  very good,  ($\psi$ shows  a  very deep
minimum)  occurs however  at a  temperature  and mass  values too  low
compared to those determined by  KA. Nevertheless, for case {\bf B} we
find also a  very deep minimum at an  effective temperature compatible
to that of KA  and a mass value almost identical to  that given in KA.
If  we accept  that this  coincidence  is not  by chance,  we have  to
conclude that this  is the best fit to  the structure and evolutionary
conditions for G117B-15A.  Thus, our results strongly suggest that the
hydrogen  envelope   present  in  G117B-15A   is  a  thick   one  with
\lmh$\approx$ -3.8.

The  result  regarding  the  actual  amount  of  hydrogen  present  in
G117B-15A should  be considered  as a very  interesting one.   In fact
this is a {\it very thick}  hydrogen envelope. Notice that this is the
upper limit we considered for  this quantity. Because of this fact, in
improving our fit  we may have considered even  thicker envelopes (see
Figs. 4-6).  However, notice  that performing asteroseismology in the
way we  have done, it is not  clear if such a  procedure would produce
stellar models plausible  from a physical point of  view.  Notice that
even  in  the hypothetical  case  that  our  artificial procedure  for
generating the stellar  models were able to handle  such models, there
is no  way to  be confident that  {\it previous} nuclear  burning were
effective  enough  to prevent  the  existence  of  such structures  in
Nature.   Because of  this reason  we think  it would  be a  more safe
procedure to compute the structure of objects of the mass of G117B-15A
starting from  the hydrogen burning  main sequence throughout  all the
evolutionary stages previous to the WD stage.  This is the only way we
envisage  in  order  to  assess  the possible  existence  of  hydrogen
envelopes thicker than those we have considered here.  In any case, we
should point out that  previous detailed calculations of the evolution
of objects  that finally form a  WD predict values  near ${\rm M_{H}}=
10^{-4} M_{\odot}$ for 0.60 \msun  objects at the start of the cooling
branch.  However  a serious study of  this problem would  carry us too
far afield and is beyond the scope of the present paper.

Finally, we can revisit the estimation of the parallax of G117B-15A as
done in  Bradley (1998). Because  the effective temperature  is higher
(11790K vs.  11600K) and the stellar  mass lower (0.52  \msun vs. 0.60
\msun,  and  thus a  larger  radius),  the luminosity  (\ll_lsun=-2.44
vs. \ll_lsun=-2.52) is higher than predicted by Bradley (1998) and for
the  same  visual apparent  magnitude  $m_{V}=  11.50$ and  bolometric
correction  of -0.611 mag  (Bergeron, Wesemael  \& Beauchamp  1995) we
find  that G117B-15A  should  have  a parallax  of  15.89 mas  whereas
Bradley  determined it  to be  16.5,  which is  still too  high to  be
compatible  (to   a  $1\sigma$   level)  with  the   determination  of
$10.5\pm4.2$ mas  by van Altena, Lee  \& Hoffleit (1994).   In view of
the amount of compatible evidence on the value of the stellar mass and
effective temperature we expect  the observed parallax of G117B-15A to
be underestimated. It seems difficult to adjudicate the discrepancy in
parallax  as due  to an  overestimation of  the value  of  the stellar
mass. Were this the case,  it should be significantly lower than found
in this work.

\begin{center}
\centerline{Table 2. {\small Summary of results for the best fitting}}
\begin{tabular}{lcc}
\hline
\hline
Quantity & Fitted model & G117-B15A \\ 
\hline
$M_*/M_{\odot}$         & 0.525  & 0.53 $^{(1)}$              \\
$\log g$                & 7.85   & 7.86 $\pm$ 0.14 $^{(1)}$   \\
$T_{\rm eff}$ [K]       & 11,790 & 11,900 $\pm$ 140 $^{(1)}$  \\
$\log(M_{\rm H}/M_*)$   & -3.83  &  $\ldots$                  \\
$\log(M_{\rm He}/M_*)$  & -2.00  &  $\ldots$                  \\
Parallax [mas]          & 15.89  & 10.5 $\pm$ 4.2 $^{(2)}$    \\
\hline
$P(k= 2)$ [s]           & 208.80       & 215.2 $^{(3)}$   \\
$P(k= 3)$               & 278.85       & 271.0 $^{(3)}$   \\
$P(k= 4)$               & 308.57       & 304.4 $^{(3)}$   \\
\hline
$\dot P(k= 2)$ [$10^{-15}{\rm s\ s}^{-1}$]& 4.43  & $2.3 \pm 1.4$ $^{(4)}$  \\
$\dot P(k= 3)$                            & 3.22  & $\ldots$      \\
$\dot P(k= 4)$                            & 5.76  & $\ldots$      \\
\hline
\end{tabular}

{\footnotesize  References:  (1) Koester  \&  Allard  (2000), (2)  van
Altena  et al.   (1994),  (3) Kepler  et  al.  (1982),  (4) Kepler  et
al. (2000)}
\end{center}

\section{Conclusions} \label{sec:conclus}

In this  paper we  have studied the  structural characteristic  of the
variable  DA white  dwarf (WD)  G117B-15A by  applying the  methods of
asteroseismology.  In  doing  so  we  have constructed  models  of  WD
evolution considering  updated and detailed  physical ingredients.  It
should be remarked than we have included several processes responsible
for the diffusion  of elements in the WD  interior.  In particular, we
considered  gravitational settling,  chemical  and thermal  diffusion.
Starting  from  an  artificial   model,  we  have  considered  several
thickness  for the  outermost hydrogen  layer, whereas  for  the inner
helium-,  carbon- and  oxygen-rich  layer we  considered the  profiles
predicted by  Salaris et al.  (1997).  The range of  stellar masses we
have considered is $0.50 \leq M_{*}/M_{\odot} \leq 0.60$. As far as we
are aware, this is the first  study in which evolutionary models of DA
WDs considering element  diffusion have been constructed for  a set of
values of hydrogen mass fraction.

The evolution of each model  sequences were followed down to effective
temperature  of 12500K  from  where on,  we  considered the  evolution
coupled to the oscillations.  We considered dipolar, adiabatic g-modes
with radial order  $k=1, \cdots, 4$.  After constructing  the full set
of  models, we considered  the location  of the  minima of  a function
conveniently  defined for the  purpose of  fitting. We  considered two
possible  mode  identifications  according  to  Bradley  (1998):  that
observed  modes 215.2,  271 and  304.4 s  correspond to  $k=1,2,3$ and
$k=2,3,4$.  We  find that the employment of  asteroseismology does not
provide an  univocal answer about the correct  identification of modes
present  in G117B-15A.  However, if  we  use them  together with  data
deduced from spectroscopy  it is found that the  identification of the
observed periods  of 215.2,  271 and 304.4  s with dipolar  modes with
$k=2,3,4$ is clearly the best.

The results  presented in this  work (summarised in table  2) strongly
suggest that G117B-15A is a DA  WD with a stellar mass of 0.525 \msun,
a  hydrogen   mass  fraction  \lmh=-3.83   and  effective  temperature
\teff$\approx$11800K. Notably, the values of the effective temperature
and stellar  mass are in very  nice agreement with  those predicted by
spectroscopic analysis  by Koester  \& Allard (2000).  This represents
the main result of the present paper.

While  the favoured value  for the  mass fraction  of hydrogen  is the
maximum we  have considered, we  have not tried  to study the  case of
slightly  (25\% say)  thicker hydrogen  layers because  our artificial
starting evolutionary technique prevents  us from being confident with
the  possibility that  such structures  can be  actually be  formed in
Nature.   The only  way to  answer this  question is  to  perform full
evolutionary  calculations  from   the  initial  stages  of  evolution
(hydrogen main sequence) to the WD stage.

Thus, performing full evolutionary studies coupled to oscillations is,
in  principle,  interesting in  two  senses.   First,  it can  provide
significative improvements to the  fitting of the oscillatory modes we
have found for G117B-15A in this paper.  Second, and more importantly,
it can  provide us with  a stringent test  of the quality  of detailed
theoretical  evolutionary   models  of  WD  stars   by  comparing  the
theoretical  pattern  of   non-radial  oscillations  against  accurate
observational determinations  of the oscillatory pattern  of WD stars.
Work in this sense is in progress and will be published elsewhere.

\section{acknowledgments} O.G.B. warmly thanks R. E. Martinez 
and E. Suarez 
for their help in getting some  key information for the
present paper.


%

\begin{figure*}
\epsfysize=600pt
\epsfbox{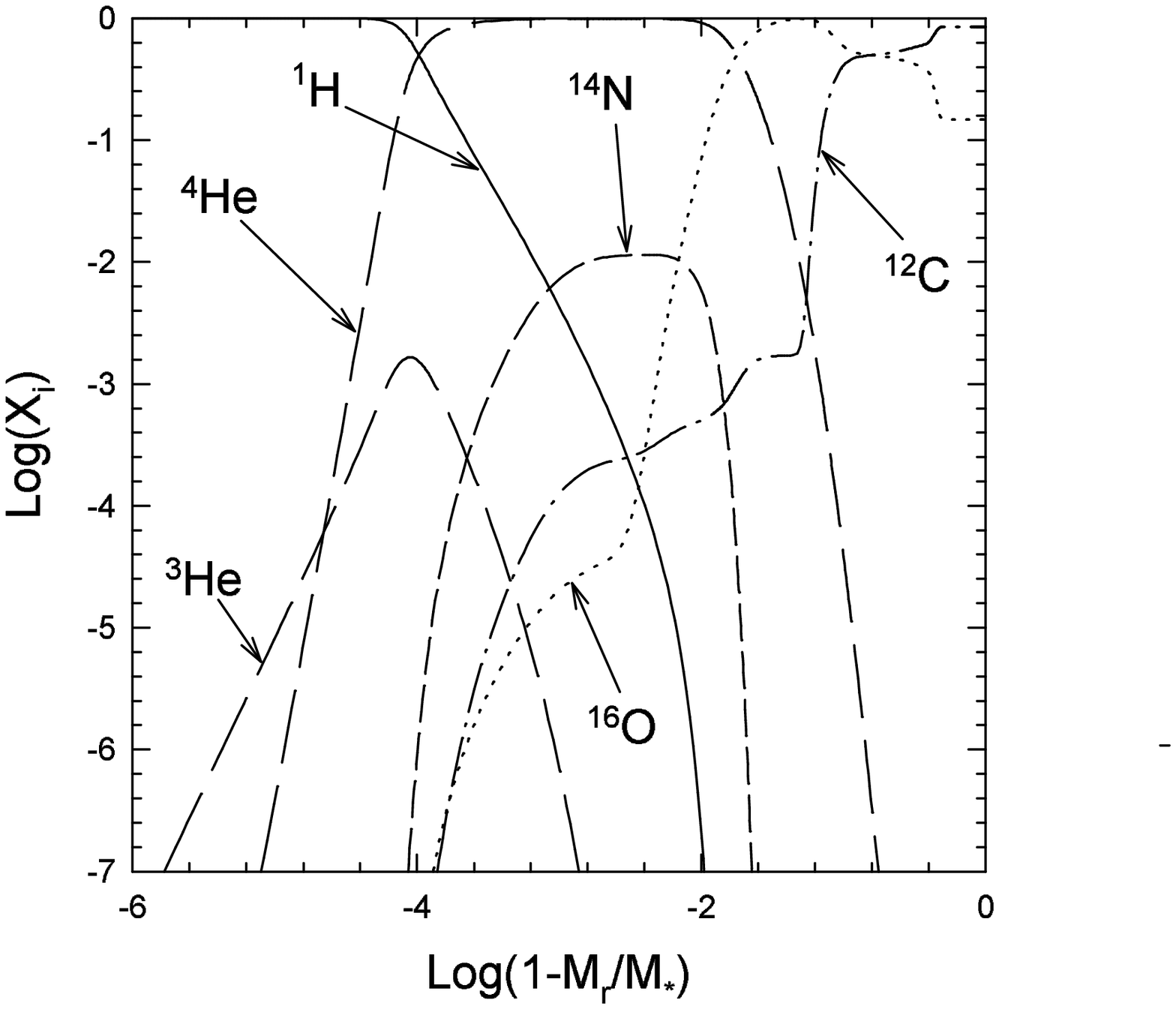} 
\caption{The internal chemical profile of a
0.55  \msun\  WD  model  for $^{1}$H,  $^{3}$He,  $^{4}$He,  $^{12}$C,
$^{14}$N, and  $^{16}$O, for the case of  \lmh=-3.862 at \lteff=4.096.
For more details, see text.} 
\end{figure*}

\begin{figure*}
\epsfysize=600pt
\epsfbox{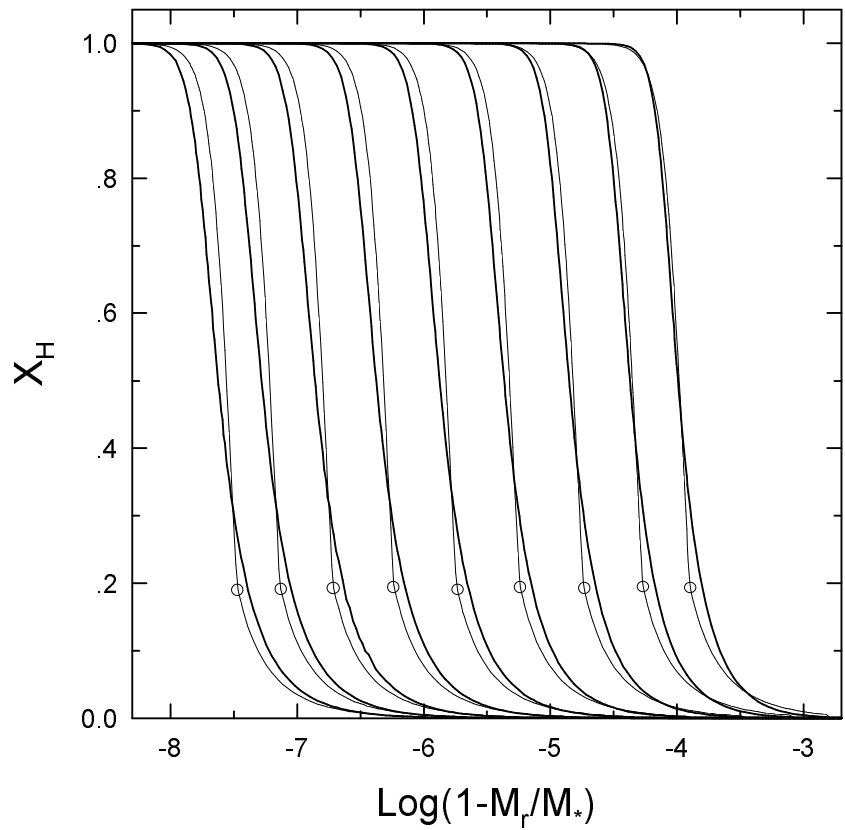} 
\caption{The profile  of the bottom  of the hydrogen envelope  for our
suite  of 0.55  \msun  models at  \teff=12,500K  represented by  solid
lines. For the  corresponding values of \mh, see  Table 1.  Thin lines
represent  the profiles  corresponding  to the  standard treatment  of
equilibrium diffusion in the  trace element approximation for the same
\mh  values.  Notice  that,  while for  thick  hydrogen envelopes  the
profiles  are rather similar,  there appear  large differences  in the
case  of thin envelopes.  Circles denote  the change  of slope  in the
profiles calculated with equilibrium diffusion.  For more details, see
text.}
\end{figure*}

\begin{figure*}
\epsfysize=600pt
\epsfbox{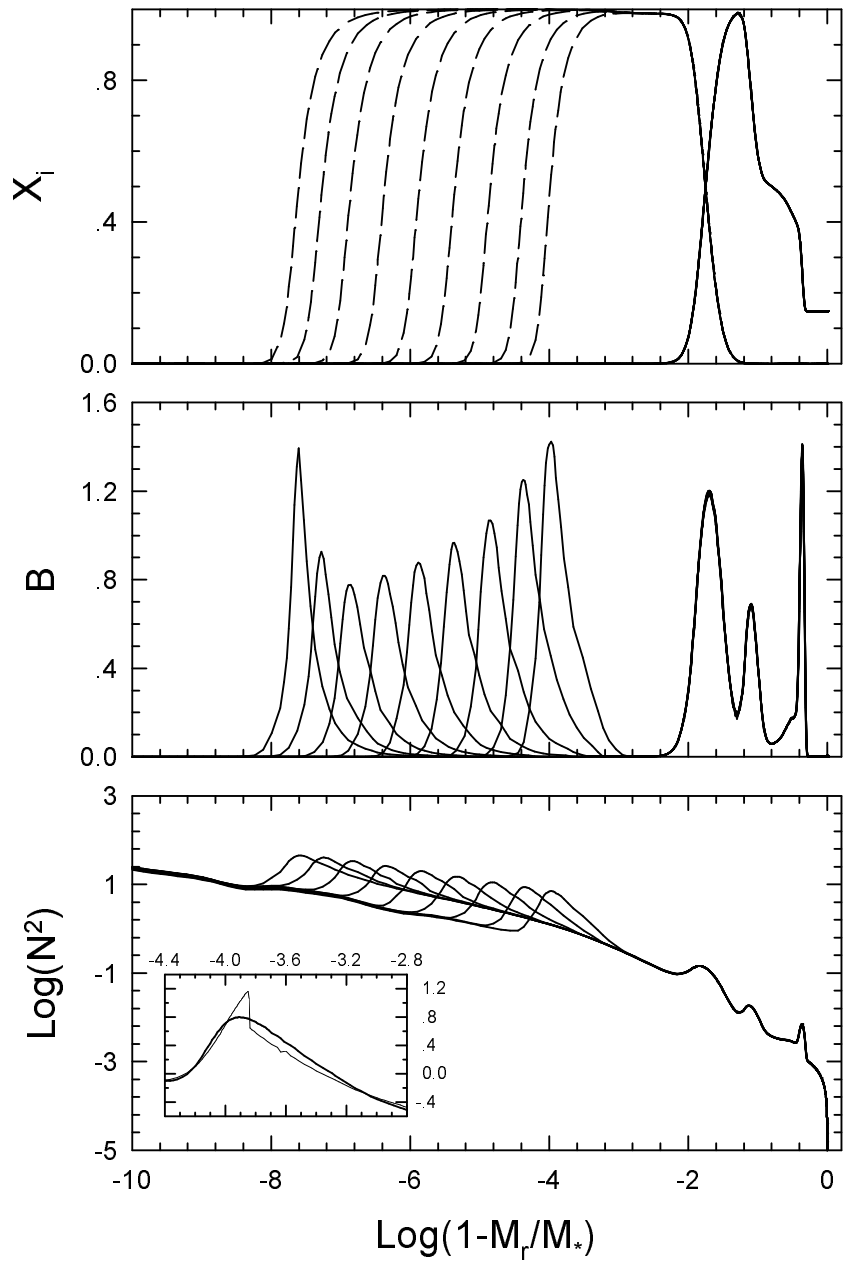} 
\caption{Some of the  main characteristics of our models  for the case
of a mass of 0.55 \msun.   {\bf Upper panel:} carbon (solid lines) and
helium (long  dashed lines) profiles.  {\bf Middle  panel:} The Ledoux
term  of the  Brunt-V\"ais\"al\"a  frequency. {\bf  Lower panel:}  The
logarithm  of  the squared  Brunt-V\"ais\"al\"a  frequency.  From  the
centre outwards,  the three peaks  of the Ledoux  term $B$ are  due to
steep slopes  in the  carbon profile.   The first two  are due  to the
structure of the carbon - oxygen interface whereas the other is due to
the helium  - carbon  interface.  The inset  in lower panel  shows the
behaviour  of the Brunt-V\"ais\"al\"a  frequency in  the case  of time
dependent diffusion  and according  to the predictions  of equilibrium
diffusion for our sequence with  the thickest hydrogen envelope at the
hydrogen - helium interface (thick and thin lines, respectively).  For
more details, see text.}
\end{figure*}

\begin{figure*}
\epsfysize=600pt
\epsfbox{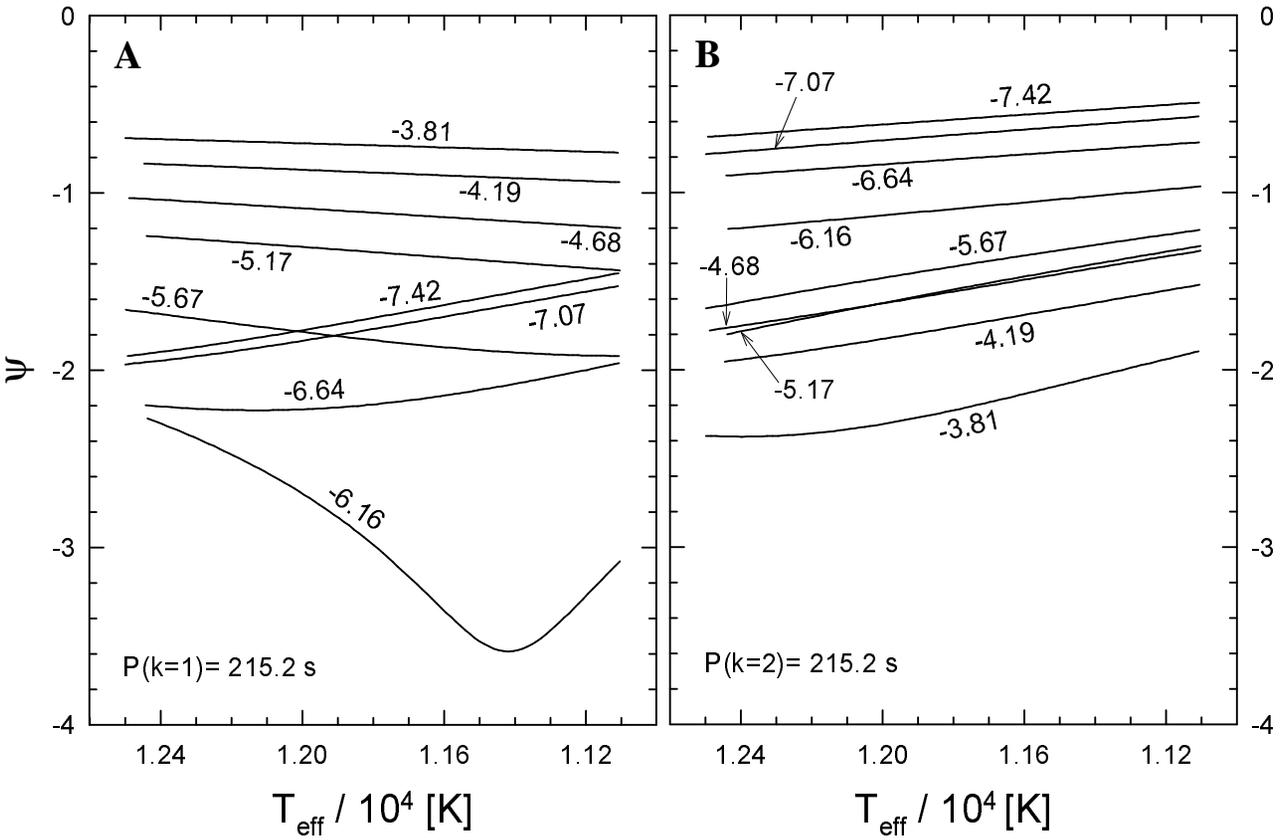} 
\caption{The value of $\psi$ as a function
of the effective temperature of the star for the case of 0.50 \msun WD
models. Curves are labelled with the respective value of \lmh. {\bf A}
and {\bf  B} stand for the  identification of the  observed modes with
$k=1,2,3$ and $k=2,3,4$ respectively.  Notice the very deep minimum of
$\psi$ for case {\bf A} at \teff=11,400 K.} 
\end{figure*}

\begin{figure*}
\epsfysize=600pt
\epsfbox{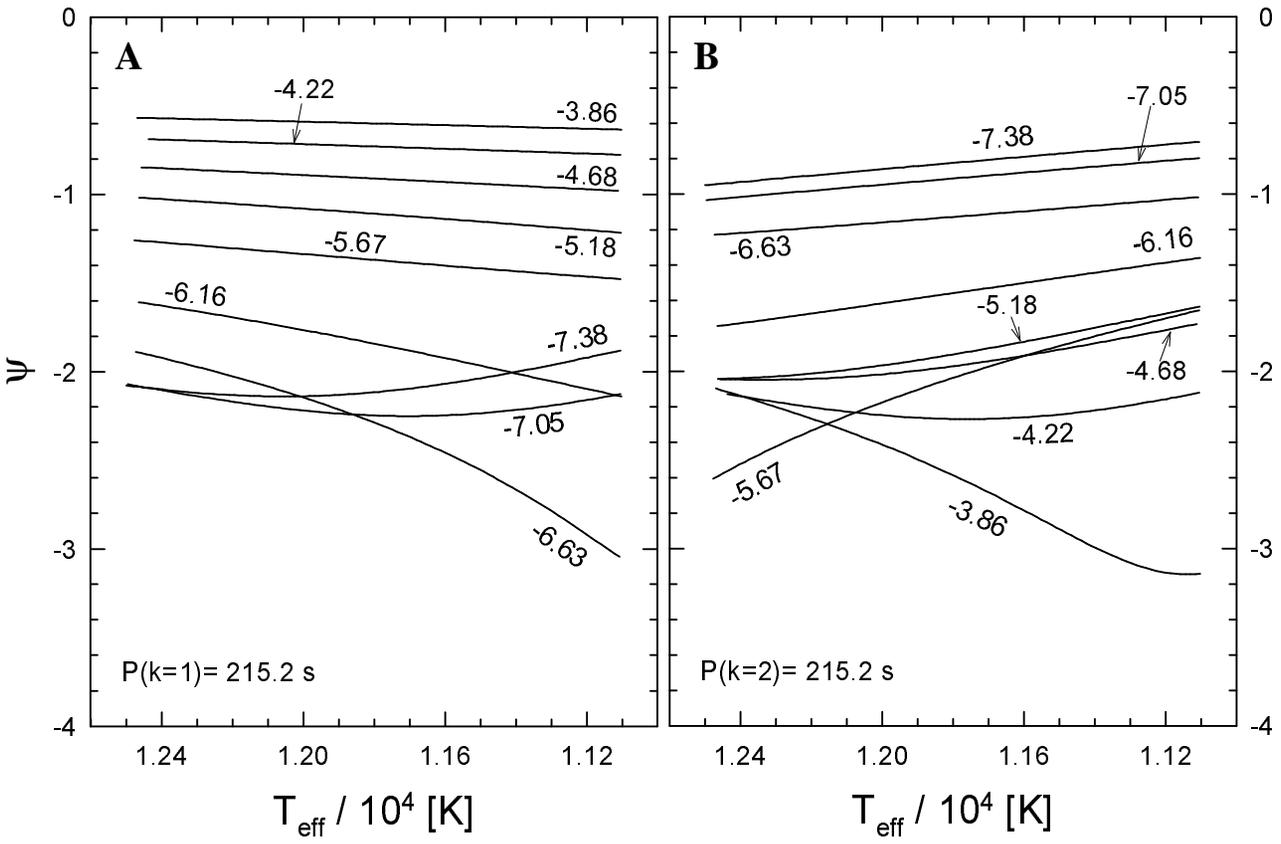} 
\caption{Same as Fig. 4 but for the case of
0.55  \msun WD models.   For this  value of  the stellar  mass, minima
corresponding to the two mode identifications have moved to the edges
of the considered interval in effective temperatures.} 
\end{figure*}

\begin{figure*}
\epsfysize=600pt
\epsfbox{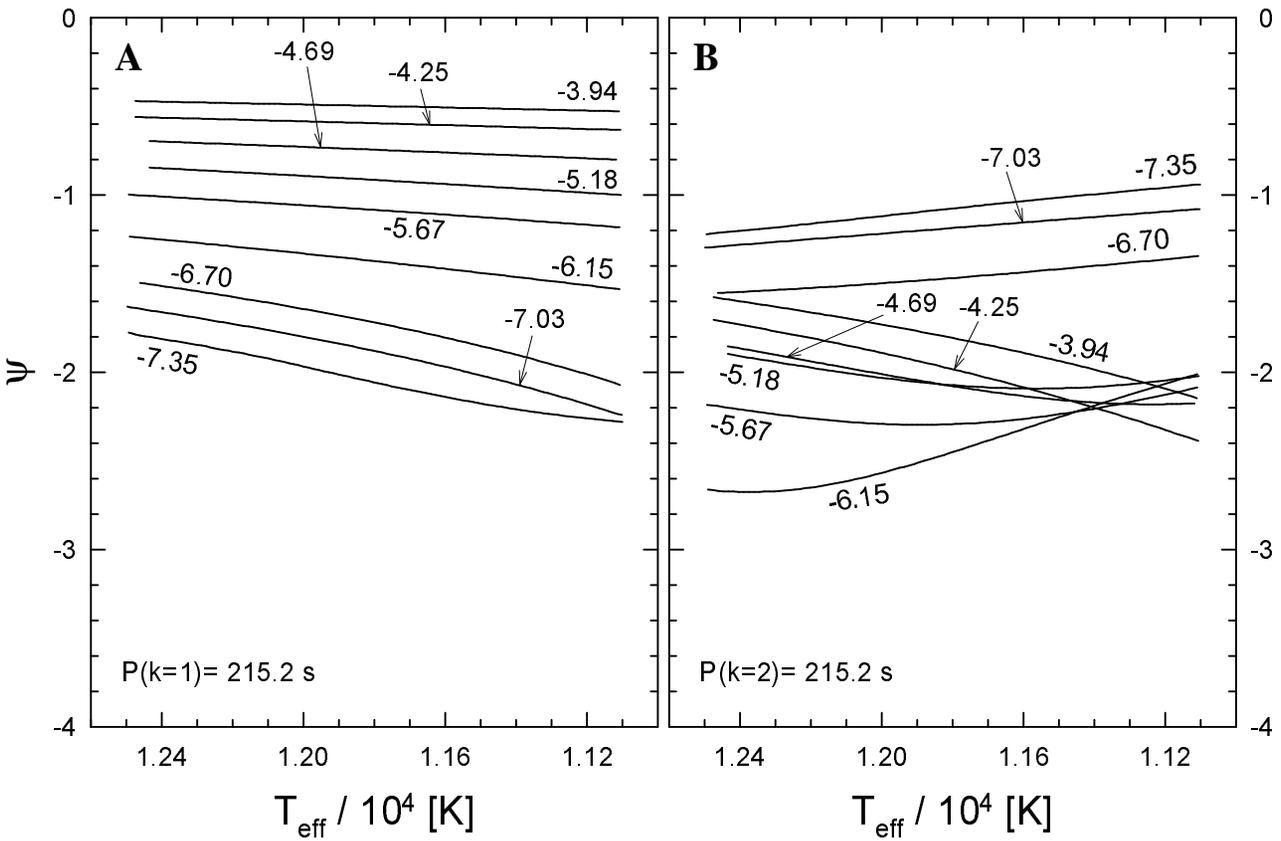} 
\caption{Same as Fig. 4  but for the case
of  0.60 \msun  WD models.   For this  value of  the stellar  mass the
values  of $\psi$  are  larger  than those  found  for the  previously
considered masses.  Thus, the mass  of G117B-15A should be  lower than
0.60 \msun.} 
\end{figure*}

\begin{figure*}
\epsfysize=600pt
\epsfbox{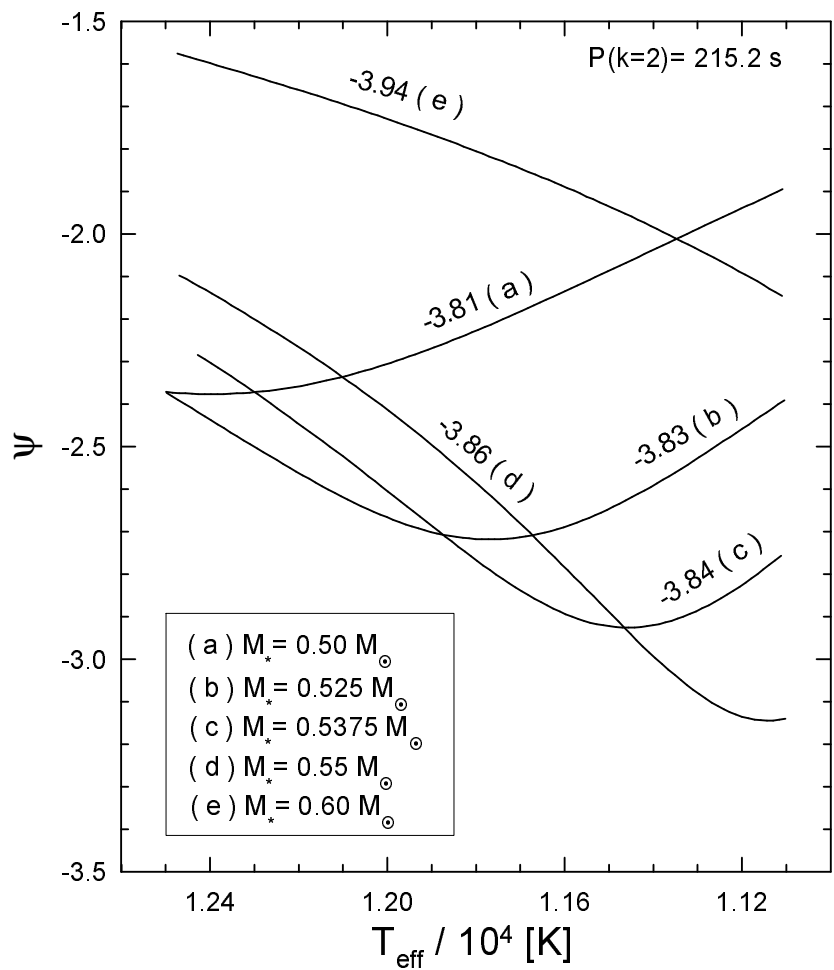} 
\caption{The value of $\psi$ as a function
of the effective temperature of the star assuming case {\bf B} for the
largest  values of  \lmh.   Notice  that if  we  assume the  effective
temperature  to be in  the range  of the  determination of  Koester \&
Allard (2000),  then, we  naturally find that  the mass of  the object
should  be 0.525  \msun \  and its  hydrogen mass  fraction  should be
\lmh=-3.83.} 
\end{figure*}

\begin{figure*}
\epsfysize=600pt
\epsfbox{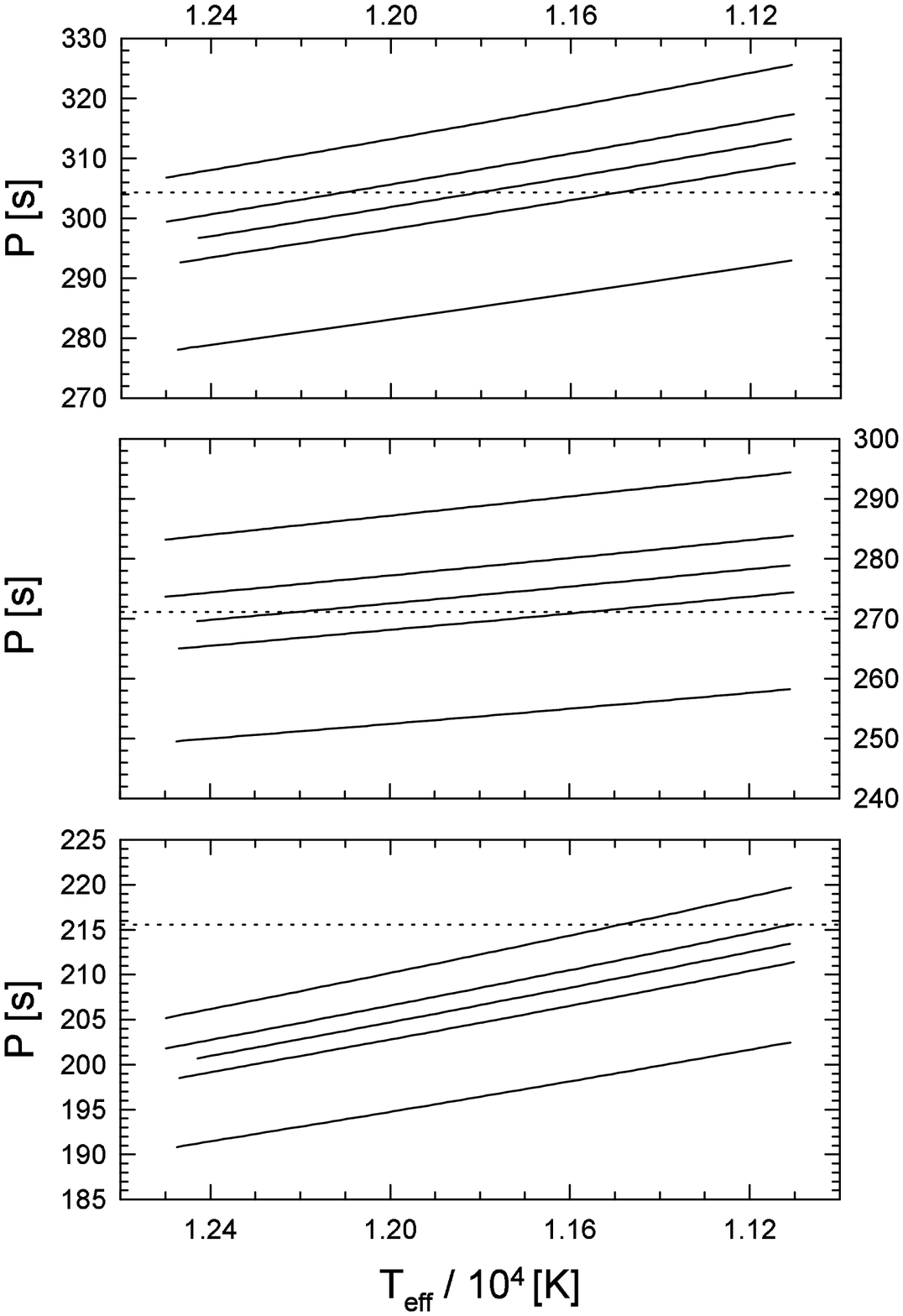} 
\caption{The periods of  oscillation for
radial order $k=2$ (lower panel) $k=3$ (middle panel) and $k=4$ (upper
panel) for the case of  models with the thickest hydrogen envelope for
each considered value of the  stellar mass (from bottom to top, curves
correspond  to  stellar masses  of  0.60,  0.55,  0.5375, 0.525,  0.50
\msun).  Notice that  the fitting to the observed  modes (denoted with
horizontal short dashed lines)  should be considered simultaneously at
the same effective temperature.} 
\end{figure*}

\label{lastpage}


\begin{thebibliography}{99}

\bibitem{} Althaus L. G., Benvenuto O. G., 2000, MNRAS, 317, 952 

\bibitem{} Althaus  L.  G., Serenelli  A. M., Benvenuto O.   G., 2001,
MNRAS, 323, 471 

\bibitem{}  Bergeron P., Wesemael  F., Beauchamp  A., 1995,  PASP, 107,
1047 

\bibitem{}  Bergeron  P., Wesemael  F.,  Lamontagne  R., Fontaine  G.,
Saffer R. A., Allard N. F. 1995, ApJ, 449, 258  

\bibitem{} Bradley P. A., 1996, ApJ, 468, 350 

\bibitem{} Bradley P. A., 1998, ApJS, 116, 307 

\bibitem{} Bradley P. A., 2001, ApJ, 552, 326 

\bibitem{} Bradley P. A., Kleinman S., 1997, in White Dwarfs, Proc. of the
10 th European workshop on white dwarfs, eds., J. Isern, E. Garc\'{\i}a-Berro
and M. Hernanz, p. 445

\bibitem{} Bradley P. A., Winget D. E., 1994, ApJ, 430, 850 

\bibitem{} Brassard P., Fontaine G.,  Wesemael F., Hansen C. J., 1992b,
ApJS, 80, 369 

\bibitem{}  Brassard P.,  Fontaine  G.,  Wesemael F.,   Talon A.,
1993,  in  White Dwarfs:  Advances  in  Observations  and Theory,  Ed.
M. A. Barstow (Dordretch, Kluwer), 485 

\bibitem{} Brassard  P., Fontaine G., Wesemael F.,  Tassoul M., 1992a,
ApJS, 81, 747 

\bibitem{}  Brassard P.,  Fontaine G.,  Wesemael F.,  Kawaler  S.  D.,
Tassoul M., 1991, ApJ, 367, 601 

\bibitem{} Brown T. M., Gilliland R. L., 1994, ARA\&A, 32, 37

\bibitem{}  Clayton D. D.,  1968, Principles  of Stellar  Evolution and
Nucleosynthesis, Mc Graw Hill 

\bibitem{} Clemens J. C., 1994, PhD Thesis, Univ. of Texas 

\bibitem{}  C\'orsico A.   H., Benvenuto  O.   G., 2002,  Ap\&SS, to  be
published (astro-ph/0104267) 

\bibitem{} C\'orsico A. H., Benvenuto  O. G., Althaus L. G., Serenelli
A. M., 2001b, MNRAS, to be published 

\bibitem{} C\'orsico A. H., Benvenuto  O. G., Althaus L. G., Isern J.,
Garc\'{\i}a - Berro E., 2001a, New Astronomy, 6, 197

\bibitem{} Dehner B. T., Kawaler S. D., 1995, ApJ, 445, L141

\bibitem{} Dolez N., Vauclair G., 1981, A\&A, 102, 375

\bibitem{} Gautschy A., Ludwig H.-G., Freytag B., 1996, A\&A, 311, 493

\bibitem{} Gautschy A., Saio H., 1995, ARA\&A, 33, 75

\bibitem{} Gautschy A., Saio H., 1996, ARA\&A, 34, 551

\bibitem{} Fontaine G.,  Brassard  P.,  1994, In IAU Colloq. 147, The
Equation of  State in Astrophysics,  Ed.  Chabrier G., Schatzman
E. (Cambridge: Cambridge Univ. Press), 347 

\bibitem{} Fontaine G., Brassard P., Wesemael F., 1994, ApJ, 428, L61 

\bibitem{} Fontaine G., Brassard P., Bergeron P., Wesemael F., 1992,
ApJ, 399, L91 

\bibitem{} Iben I. Jr., MacDonald J., 1985, ApJ, 296, 540  

\bibitem{} Iben I. Jr., MacDonald J., 1986, ApJ, 301, 164  

\bibitem{} Kepler  S.\ O., Robinson E. L., Nather R.\ E.,
McGraw J.\ T.,  1982, ApJ, 254, 676  

\bibitem{} Kepler  S.\ O., Mukadam A.,  Winget D.\ E.,  Nather R.\ E.,
Metcalfe T.\ S.,  Reed M.\ D., Kawaler S.\ D.,  Bradley P.\ A.,\ 2000,
ApJ, 534, L185  

\bibitem{} Koester D., Allard N. F., 2000, Baltic Astron., 9, 119 

\bibitem{} Metcalfe T.\  S., Nather R.\ E., Winget  D.\ E., 2000, ApJ,
545, 974  

\bibitem{} Metcalfe T.\ S., Winget D.\ E., Charbonneau P.,  2001, ApJ,
557, 1021 

\bibitem{} MacDonald J., Hernanz M., Jose J., 1998, MNRAS, 296, 523  

\bibitem{} Mc Graw J. T., Robinson E. L.,  1976, ApJ, 205, L155 

\bibitem{} Montgomery M.\ H., Winget, D.\ E., 1999, ApJ, 526, 976  

\bibitem{} Montgomery M.  H., Metcalfe T.  S., Winget D. E., 2001, ApJ,
548, L53  

\bibitem{} Pfeiffer B. et al., 1996, A\&A, 314, 182

\bibitem{} Robinson  E.  L., Mailloux  T.  M., Zhang E.,  Koester D.,
Steining R.   F., Bless R.   C., Percival J.   W., Taylor M.   J., van
Citters G.  W., 1995, ApJ, 438, 908 

\bibitem{}  Salaris M.,  Dom\'{\i}nguez  I., Garc\'{\i}a-Berro  E., 
Hernanz  M.,
Isern J., Mochkovitch R.,\ 1997, ApJ, 486, 413 

\bibitem{} Tassoul M., Fontaine G., Winget D. E.,  1990, ApJS, 72, 335 

\bibitem{} Unno W.,  Osaki Y., Ando H., Saio  H., Shibahashi H., 1989,
Nonradial Oscillations of Stars, University of Tokio Press, 2nd. ed. 

\bibitem{} van  Altena W. F.,  Lee J. T.,  Hoffleit E. D.,  1994, The
General  Catalogue  of   Trigonometric  Parallaxes  (New  Haven:  Yale
Univ. Obs.) 

\bibitem{}  Van Kerkwijk  M. H.,  Bell J.  F., Kaspi  V.  M., Kulkarni
S. R., 2000, ApJ, 530, L37 

\bibitem{}    Winget   D.     E.,    1988,   in    IAU   Symp.    123,
Eds. E.  J. Christensen -  Dalsgaard, S. Frandsen  (Dordrecht Reidel),
p. 305

\bibitem{} Winget  D. E., Van  Horn H. M.,  Tassoul M., Hansen  C. J.,
Fontaine G., Carroll B. W., 1982, ApJ, 252, L65

\bibitem{}    Winget   D.     E. et al.,  1991, ApJ, 378, 326

\end{thebibliography}
\end{document}